\documentclass[aps,prl,reprint,twocolumn,superscriptaddress,showpacs]{revtex4-2}

\usepackage{amsmath, amssymb}
\usepackage{makecell}
\usepackage{multirow}
\usepackage{graphicx}
\usepackage{mathrsfs}
\usepackage{bm}
\usepackage{amsmath}
\usepackage{dcolumn}
\usepackage{epstopdf}
\usepackage{dsfont}
\usepackage{amssymb}
\usepackage{tabularx}
\usepackage{array}
\usepackage{float}
\usepackage{color}
\usepackage{epstopdf}
\usepackage{mathrsfs}
\usepackage[colorlinks, linkcolor=blue,anchorcolor=blue,citecolor=blue,urlcolor=blue]{hyperref}
\usepackage{extarrows}
\usepackage{braket}

\begin{document}
\title{Quantum Metric Nonlinear Spin-Orbit Torque Enhanced by Topological Bands}

\author{Xukun Feng}
\affiliation{Science, Mathematics and Technology, Singapore University of Technology and Design, Singapore 487372, Singapore}

\author{Weikang Wu}
\affiliation{Key Laboratory for Liquid-Solid Structural Evolution and Processing of Materials
(Ministry of Education), Shandong University, Jinan 250061, China}

\author{Hui Wang}
\affiliation{Division of Physics and Applied Physics, School of Physical and Mathematical Sciences, Nanyang Technological University, Singapore 637371, Singapore}

\author{Weibo Gao}
\affiliation{Division of Physics and Applied Physics, School of Physical and Mathematical Sciences, Nanyang Technological University, Singapore 637371, Singapore}

\author{Lay Kee Ang}
\affiliation{Science, Mathematics and Technology, Singapore University of Technology and Design, Singapore 487372, Singapore}

\author{Y. X. Zhao}
\affiliation{Department of Physics, The University of Hong Kong, Hong Kong, China}

\author{Cong Xiao}
\email{congxiao@um.edu.mo}
\affiliation{Institute of Applied Physics and Materials Engineering, University of Macau, Macau, China}

\author{Shengyuan A. Yang}
\email{yangshengyuan@um.edu.mo}
\affiliation{Institute of Applied Physics and Materials Engineering, University of Macau, Macau, China}
%\affiliation{Research Laboratory for Quantum Materials, IAPME, Faculty of Science and Technology, University of Macau, Macau, China}

\begin{abstract}
Effects manifesting quantum geometry have been a focus of physics research. Here, we reveal that quantum metric plays a crucial role in nonlinear electric spin response, leading to a quantum metric spin-orbit torque. We argue that enhanced quantum metric can occur at band (anti)crossings, so the nonlinear torque could be amplified in topological metals with nodal features close to Fermi level. By applying our theory to magnetic Kane-Mele model and monolayer CrSBr, which feature nodal lines and Weyl points, we demonstrate that the quantum metric torque dominates the response, and its magnitude is significantly enhanced by topological band structures, which even surpasses the previously reported linear torques and is sufficient to drive magnetic switching by itself.

%While the nonlinear current induced spin-orbit torque is a central topic in the emerging field of nonlinear spintronics, what kind of materials can enhance this effect and whether or not it is significant enough for magnetic switching are unknown, posing critical objectives on its study.   In this work, we unveil that topological nodal-line band structures can support an unprecedentedly strong nonlinear torque, and the two-band character of nodal-line structures forces this torque to be determined by a novel quantum metric in the extended parameter space spanned by crystal momentum and magnetization. First-principles calculations in monolayer CrSBr show a giant quantum metric torque, which may lead to magnetization reversal by nonlinear spin response for the first time at a moderate driving current density.
%This work opens the door to magnetic switching by nonlinear torque, and uncovers the role in magnetic memory played by topological nonlinear spintronics, paving the way for novel device applications of this emerging field.

\end{abstract}
\maketitle

Quantum geometry, the geometry of Hilbert space for a quantum system, plays critical roles in many physical processes.
Particularly, physics associated with Berry curvature and Berry phase have been extensively studied since 1980s \cite{berry1984quantal,Xiao2010Berry,bohm2003geometric,shapere1989geometric}. Meanwhile, quantum metric, another important ingredient of quantum geometry \cite{QM1980,Marzari1997Maximally-PRB}, was less explored for a long time. In the past few years, quantum metric started to attract significant attention, because it was found to underlie several interesting effects, such as orbital magnetic susceptibility \cite{Gao2015}, flat-band superconductivity \cite{Torma2015,Torma2016,Bernevig2020,Bernevig2022,Torma2022}, second and third order Hall effects \cite{Gao2014,Wang2021,Liu2021,Xu2023QM,Gao2023QM,Lai2021}, and etc \cite{Qi2015,ExcitonQM2015,Gao2019QMD,QMmeasurement2020,Yang2020,Ahn2020,Watanabe2021,Yang2021,Ahn2022,Justin2022,Bhalla2022}. Currently, the research focus is to explore more physical phenomena with such quantum geometry perspective, and to utilize the new understanding to advance applications \cite{Torma2023}.

Current induced spin polarization is a central effect in spintronics \cite{Manchon2019}. In a single piece of ferromagnet, the effect induces a
spin-orbit torque on the magnetization vector, which may enable electric
control of magnetization, desired for information device applications \cite{Manchon2008,Geller2009,Franz2010,Miron2010,Vyborny2011,Miron2011,Liu2012,Garello2013,Kurebayashi2014,Freimuth2014,Jungwirth2016,Yuriy2017}. However, the effect at linear order, i.e., the response linear in applied $E$ field or current, is forbidden in centrosymmetric systems. To make use of it, one has to adopt
low-symmetry magnets or fabricate heterostructures. Recently, the effect has been extended to nonlinear order, where the induced spin polarization  $\delta S$ and its spin-orbit torque scale with $E^2$ \cite{Xiao2022NLSOT,Xiao2023NLSOT}. Particularly, in magnets,
this effect has an \emph{intrinsic} contribution, determined solely by the magnetic band structure \cite{Xiao2022NLSOT},
and the signal of such intrinsic nonlinear electric spin generation was detected in a very recent experiment on Pt-Py bilayers \cite{Kodama2023}. This opens new possibility of nonlinear spintronics.
Nevertheless, the study is still in its infancy, with two outstanding questions unanswered. First, it is not clear whether the nonlinear spin-orbit torque by itself is large enough for magnetic reversal in real materials.
Second, what kind of materials can be good platforms to enhance this effect?

%In the field of spintronics, current induced spin polarization is a central effect enabling electric
%control of magnetization.
%
%
%
%Much theoretical and experimental effort \cite{Manchon2008,Geller2009,Garate2009,Franz2010,Miron2010,Vyborny2011,Miron2011,Liu2012,Garello2013,Kurebayashi2014,Freimuth2014,Jungwirth2016,Yuriy2017} has been devoted to generating spin polarization by virtue of spin-orbit coupling in the linear order of an applied electric field. The study on this subject known as spin-orbit torque has very recently been extended to the nonlinear order \cite{Xiao2022NLSOT,Xiao2023NLSOT,Kodama2023}, which can play a leading role in magnets with inversion symmetry or weak inversion breaking. This development constitutes a building block of the emerging field of nonlinear spintronics \cite{he2019}, and extends the context of nonlinear electronics \cite{ma2019observation,Kang2019,Lu2021,Xu2023QM}. Nevertheless, the study of nonlinear spin-orbit torque is still in its infancy, and two outstanding questions stand. First, what classes of materials can enhance this effect? Second, whether or not it is significant enough for magnetization reversal?

In this work, we establish a connection between quantum geometry and nonlinear spintronics, which also helps address the above questions. We unveil that the intrinsic nonlinear spin-orbit torque has a dominant contribution from quantum metrics in the extended parameter space spanned by momentum and magnetization. We argue that such a quantum metric torque (QMT) is greatly magnified at band (anti)crossings, where the interband mixing is strong and local gap is small. Based on this understanding, we propose that magnetic topological metals could be ideal systems to enhance QMT. We demonstrate our theory in a magnetic Kane-Mele model and via first-principles calculations in an existing two dimensional (2D) magnet CrSBr. For CrSBr, we find that the topological band-enhanced QMT can surpass the usual values of linear spin-orbit torque by more than one order of magnitude, and it indicates for the first time the possibility of magnetic switching by nonlinear spin response at a moderate driving current density of $10^6$ to $10^7\ \text{A/cm}^2$. Our work uncovers the significance of quantum metric in nonlinear spintronic responses, suggests topological metals as promising platforms to amplify nonlinear spin responses, and reveals the potential of QMT for designing full electrically controlled spintronic devices.

%answer these two questions. We unveil that topological nodal-line materials can offer unprecedentedly strong
%nonlinear intrinsic spin-orbit torques. Meanwhile, the two-band character of nodal-line structures forces this torque to be dictated by a novel band geometric quantity -- the Fubini-Study quantum metric in the extended parameter space spanned by crystal momentum and magnetization. Combining the theory with first-principles calculations in a real material, monolayer CrSBr, we find giant quantum metric torques that surpass the usual values of linear spin-orbit torque by more than two orders of magnitude, and show for the first time the possibility of magnetic switching by nonlinear spin response at a moderate driving current density of $10^6\sim 10^7 \text{A/cm}^2$. This work not only opens up the study of magnetic reversal by nonlinear spin-orbit torque of quantum metric origin, but also uncovers that a new research branch of topological nonlinear spintronics emerging from the combination of nonlinear spintronic effects and topological band structures can play a significant role in magnetic memory, pointing to an avenue for incorporating nonlinear spintronics into the standard application framework of magnetic devices.

\textcolor{blue}{\emph{Quantum metric \& nonlinear spin response.}} Quantum metric measures the distance between quantum states. It is the real part of quantum geometric tensor (also known as the Fubini-Study metric). For example, in momentum space, the quantum metric for a band with index $n$ takes the form of \cite{QM1980}
\begin{equation}
g_{ab}^{n}(\bm k)=\operatorname{Re}\left\langle\partial_{k_{a}} u_{n}|(1-|u_{n}\rangle\langle u_{n}|)| \partial_{k_{b}} u_{n}\right\rangle,
\end{equation}
where $|u_n\rangle$ is the Bloch eigenstate (for simple notations, the $k$ dependence is not explicitly shown in the expression), and $a$ and $b$ label Cartesian components. Using the completeness relation, the quantum metric can be expressed as a sum $g^{n}_{ab}=\sum_{\ell \neq n} g^{n\ell}_{ab}$, in terms of its band components
$
  g^{n\ell}_{ab}(\bm k)=\operatorname{Re}\left[\langle\partial_{k_{a}} u_{n} | u_{\ell}\rangle \langle u_{\ell} | \partial_{k_{b}} u_{n}\rangle\right].
$

Importantly, $g^{n\ell}_{ab}$ also appears in the momentum-space Berry connection polarizability (BCP) tensor $G_{ab}$ \cite{Gao2014,Liu2022}, which characterizes how the position of an electron wavepacket is shifted by applied $E$ field. Specifically,
\begin{equation}
    G_{ab}^{n}(\bm k)=2 \sum_{\ell \neq n} \frac{g_{ab}^{n\ell}}{\varepsilon_{n}-\varepsilon_{\ell}},
\end{equation}
where $\varepsilon_{n}$ is the band energy, and we set $e=\hbar=1$.

\begin{figure}
	\includegraphics[width=8.6cm]{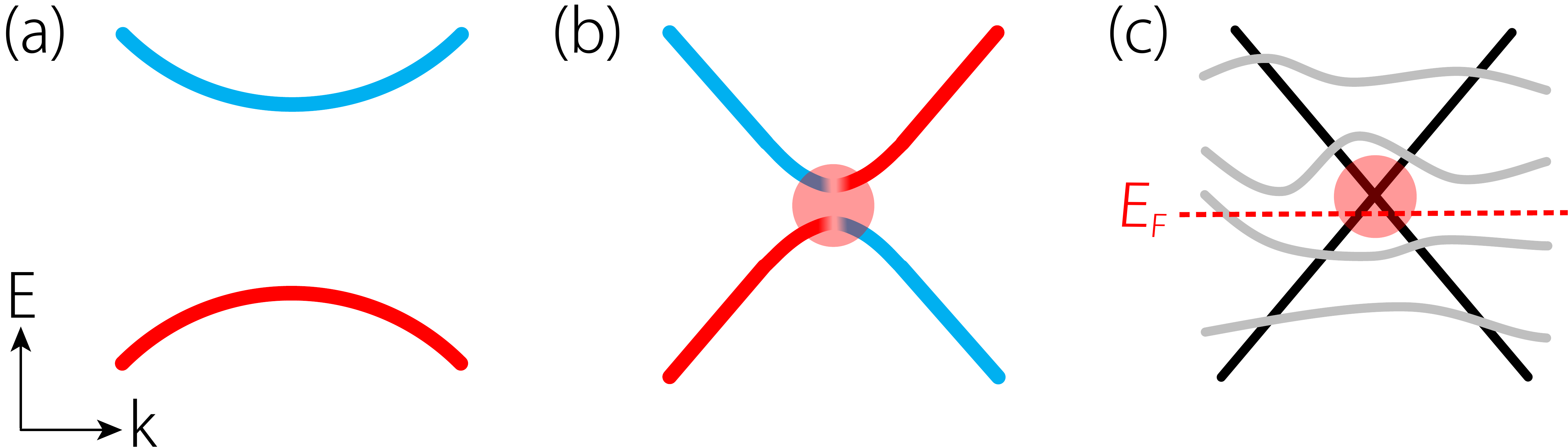}
	\caption{(a) Well separated bands with relatively small quantum metric. The blue and red colors here represent
different character of the states. (b) Quantum metric is pronounced at a band (anti-)crossing.
(c) Such band (anti-)crossings around Fermi level act like hot spots for large quantum metric in a general band structure.
		\label{fig1}}
\end{figure}

What kind of band feature tends to give a large quantum metric and BCP? First, let us consider the simple case of a two-band model. We have
\begin{equation}
  G_{ab}^{n}(\bm k)= \frac{2 g_{ab}^{n}}{\varepsilon_{n}-\varepsilon_{\bar{n}}},
\end{equation}
with $n,\bar{n}\in \{1,2\}$ and $\bar{n}\neq n$. Usually, different bands originate from different atomic orbitals. When the two bands are well separated, as illustrated in Fig.\ref{fig1}(a), one expects that the quantum metric $g^n$ is relatively small for each band, since a state is similar to its neighbors in a band. $g^n$ should be enhanced when the two bands are close to each other and start to have strong interband mixing, especially when they form a band (anti)crossing, as illustrated in Figs. \ref{fig1}(b) and \ref{fig1}(c). Near the (anti)crossing point, a state could have dramatically different characteristics from its neighbors, corresponding to a large distance in projective Hilbert space, hence a large $g^n$. Furthermore, the small local gap $\Delta \varepsilon$ near the (anti)crossing also helps magnify $g^n$ and $G^n$, as one can show that they scale as $1/(\Delta \varepsilon)^{2}$ and $1/(\Delta \varepsilon)^{3}$, respectively.

These considerations also extend to multi-band systems. Generally, band (anti)crossing points would act as ``hot spots'' for quantum metric and BCP.  Motivated by these observations, we decompose BCP into two parts:
\begin{equation}\label{BCP}
  G_{ab}^{n}(\bm k)=\frac{2 g_{ab}^{n}}{\varepsilon_{n}-\varepsilon_{\bar{n}}}+\delta G^n,
\end{equation}
where the first term explicitly involves the quantum metric for band $n$, and $\bar{n}$ here labels the band whose energy is closest to $n$ at $\bm k$; and the second term, as the difference between $G^n$ and the quantum metric term, involves additional interband contribution from remote bands, which is usually small (its expression is given in Supplemental Material \cite{supp}). Hence, it is almost always the quantum metric term that dominates whenever $G^n$ takes significant values.

The connection between quantum metric and nonlinear torque is made via BCP and Eq.~(\ref{BCP}).
Considering a centrosymmetric magnet, the torque arises from the nonlinear spin polarization $\delta S$ induced by applied $E$ field, with $\delta S_a = \alpha_{abc}E_b E_c$ and $\alpha$ the response tensor.  Here, we focus on the intrinsic contribution, which is a band structure property independent of carrier scattering. It can be expressed as \cite{Xiao2022NLSOT}
%
%
%Particularly, recent work predicts an intrinsic nonlinear spin-orbit torque in magnets, which
%
%
%
%We first show the quantum metric nature of intrinsic nonlinear spin polarization, which is quantified by $\delta S_i = \alpha^{\mathrm{int}}_{ij\ell}E_j E_\ell$, where $i,j,\ell$ are Cartesian indices. According to the previous theory \cite{Xiao2022NLSOT}, the intrinsic nonlinear spin response coefficient $\alpha^{\mathrm{int}}_{ij\ell}$, with the last two indices symmetrized, is given by (set $e=\hbar=1$)
\begin{equation}\label{total}
\begin{aligned}
    \alpha_{abc}=-\frac{1}{2}\sum_{n\bm k}[f_{0} \partial_{m_{a}} G_{bc}^{n}+f_{0}' (s_{a}^{n} G_{bc}^{n}+v_{b}^{n} \mathfrak{G}_{ac}^{n}+v_{c}^{n} \mathfrak{G}_{ab}^{n})].
\end{aligned}
\end{equation}
Here, $f_0$ is the Fermi-Dirac distribution function, $s^n$ and $v^n$ denote the expectation values of spin and velocity operators in state $|u_n\rangle$, and $\bm m$ is the magnetization. Besides the momentum-space BCP, there also appears the BCP $\mathfrak{G}$ in magnetization space. Analogous to Eq.~(\ref{BCP}), we may write
\begin{equation}\label{km}
  \mathfrak{G}_{ab}^{n}(\bm k)=\frac{2 \mathfrak{g}_{ab}^{n}}{\varepsilon_{n}-\varepsilon_{\bar{n}}}+\delta \mathfrak{G}^n,
\end{equation}
where
\begin{equation}
    \mathfrak{g}_{ab}^{n}(\bm k)=\operatorname{Re}\langle\partial_{m_{a}} u_{n}|(1-|u_{n}\rangle\langle u_{n}|)| \partial_{k_{b}} u_{n}\rangle
\end{equation}
is the quantum metric in the $m$-$k$ space, and again this metric term typically dominates over the $\delta \mathfrak{G}^n$ term.

Keeping only the quantum metric terms in (\ref{BCP}) and (\ref{km}), we obtain the quantum metric contribution:
\begin{equation}\label{FSM}
\begin{aligned}
    \alpha^\text{QM}_{abc}=-\sum_{n\bm k} \Bigg(f_{0} \partial_{m_{a}} \frac{g_{bc}^{n}}{\varepsilon_{n}-\varepsilon_{\bar{n}}}+f_{0}^{\prime} \frac{s_{a}^{n} g_{bc}^{n}+v_{b}^{n} \mathfrak{g}_{ac}^{n}+ v_{c}^{n} \mathfrak{g}_{ab}^{n}}{\varepsilon_{n}-\varepsilon_{\bar{n}}}\Bigg).
\end{aligned}
\end{equation}
The key point is that $\alpha^\text{QM}_{abc}$ gives the dominant contribution to the intrinsic response. As we shall demonstrate later, in fact, one usually has
\begin{equation}\label{dom}
  \alpha_{abc}\approx \alpha^\text{QM}_{abc},
\end{equation}
when the response is appreciable.
Therefore, the intrinsic nonlinear spin polarization and its resulting spin-orbit torque manifest the quantum metrics $g^n$ and $\mathfrak{g}^n$ for the extended parameter space spanned by momentum and magnetization. Detecting these signals offers us a new route to probe quantum metrics.

With this new perspective, to enhance the nonlinear torque, it is natural to consider ferromagnetic metals with band (anti)crossings around Fermi level, which can contribute large quantum metrics. We shall demonstrate this idea in the following discussion.

\begin{figure}
	\includegraphics[width=8.6cm]{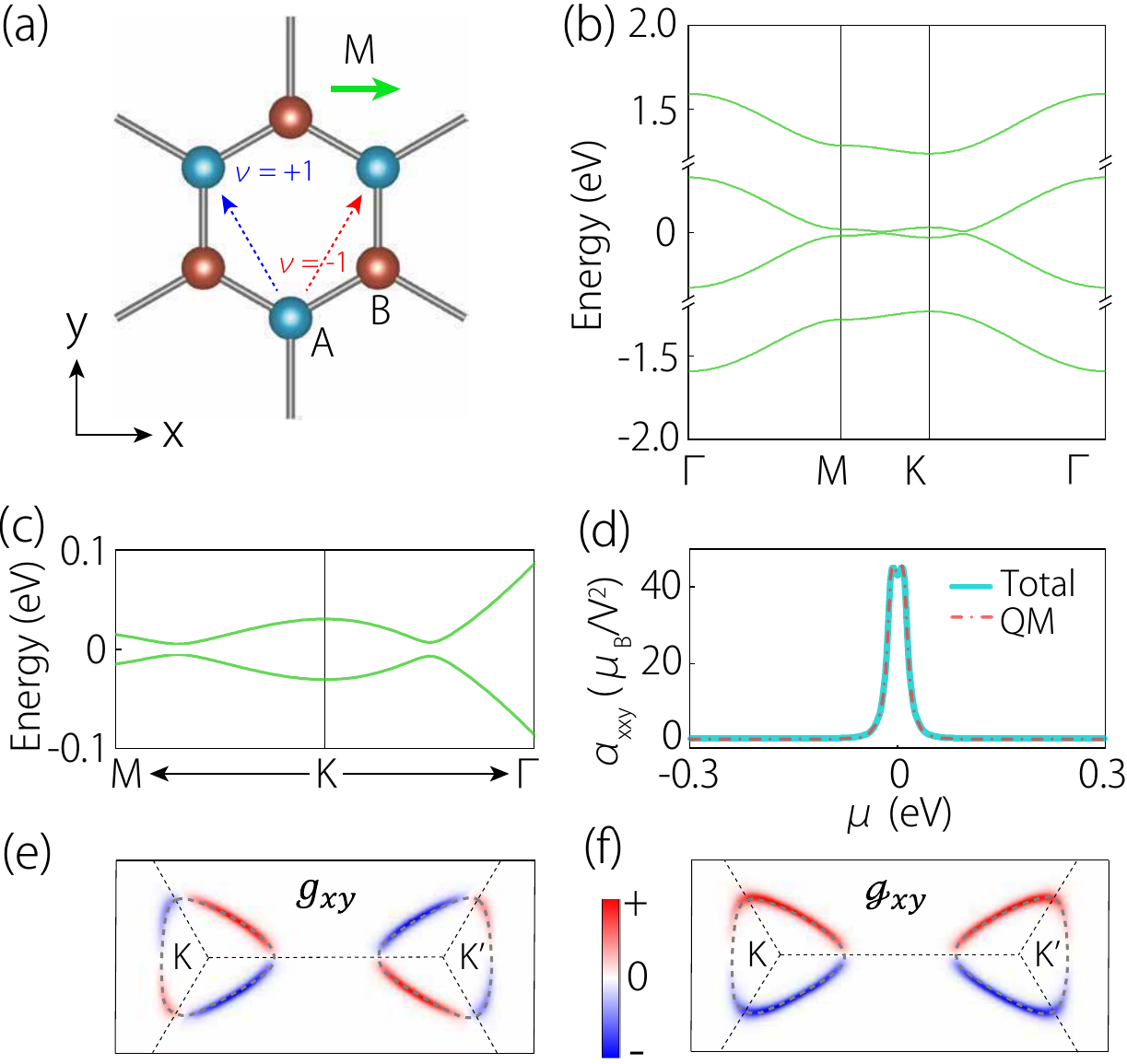}
	\caption{(a) Top view of the honeycomb lattice for the magnetic Kane-Mele model. The green arrow represents the direction of in-plane exchange field. (b) Typical band structure of the model.  (c) Enlarged view of the low-energy band structure around $K$, showing a gapped nodal loop. (d) Calculated total intrinsic nonlinear spin response $\alpha_{xxy}$ and the quantum metric contribution in this response versus chemical potential. (e) and (f) show the distributions of quantum metric $g_{xy}$ and $\mathfrak{g}_{xy}$ of the top valence band in the Brillouin zone, which are concentrated around the two gaped nodal loops as indicated by the dashed lines. In the model, we take $t = 0.25$ eV, ${t}_\text{SO}$ = 0.005 eV, $\Delta_M$ = 1.26 eV, and $\lambda_v$ = 0.6 eV.
		\label{fig2}}
\end{figure}

\textcolor{blue}{\emph{Magnetic Kane-Mele model.}} To illustrate the features of the quantum metric nonlinear response, we study the ferromagnetic Kane-Mele model defined on a honeycomb lattice [Fig.~\ref{fig2}(a)] \cite{kane2005z,chen2020universal}. Its Hamiltonian can be written as
\begin{equation}
\begin{aligned}
    H = &~t\sum_{\langle i j\rangle, \alpha}  c_{i \alpha}^{\dagger} c_{j \alpha} + i t_\text{SO}\sum_{\langle\langle i j\rangle\rangle, \alpha \beta}  \nu_{i j} \sigma_{\alpha \beta}^{z} c_{i \alpha}^{\dagger} c_{j \beta} \\
    & + \frac{\Delta_{M}}{2}\sum_{i}(\hat{\boldsymbol{m}} \cdot \boldsymbol{\sigma})_{\alpha \beta} c_{i \alpha}^{\dagger} c_{i \beta} + \lambda_{v} \sum_{i, \alpha} \xi c_{i \alpha}^{\dagger} c_{i \alpha}.
\end{aligned}
\end{equation}
Here, $c_{i \alpha}^{\dagger}$ ($c_{i \alpha}$) is creation (annihilation) operator for an electron at site $i$ with spin $\alpha$, the first term is the nearest-neighbor hopping, the second term is the intrinsic spin-orbit coupling for second neighbor hopping, with $\nu_{ij} = +1(-1)$ when the electron makes a right (left) turn in the hopping process, the third term accounts for the ferromagnetic exchange, with $\hat{\bm m}$ the unit vector for magnetization direction, and the last term with $\xi=\pm 1$ represents the staggered sublattice potential.

%$c_{i \alpha}$ ($c_{i \alpha}^{\dagger}$) represents the annihilation (creation) operator for an electron with spin $\sigma$ at site $i$. The unit vector $\hat{\boldsymbol{n}}$ represents the direction of the exchange field. The first term in this model corresponds to the nearest-neighbor hopping, while the second term represents the next-nearest-neighbor hopping, signifying the intrinsic spin-orbit coupling with $\nu_{ij} = +1(-1)$ as defined in Fig.~\ref{fig1}(a). The third term accounts for the exchange field, specifically an exchange field along the x-direction, as depicted in Fig.~\ref{fig1}(a). The fourth term represents the staggered sublattice potential, where $\xi = +1(-1)$ is defined for the sublattice A(B).

A typical band structure of this model is shown in Fig.~\ref{fig2}(b), where we take $\hat{\bm m}$ to be along the $x$ direction. One observes that two bands almost touch around zero energy. In fact, without magnetization ($\Delta_M=0$), the system is a magnetic nodal-loop semimetal, with the two low-energy bands crossing at two nodal loops centered at $K$ and $K'$ points, which are protected by the horizontal mirror $\mathcal{M}_z$. The magnetism breaks $\mathcal{M}_z$ symmetry and opens a small gap at the nodal loops (see Figs.~\ref{fig2}(b) and \ref{fig2}(c)).

%only the two central bands are in close proximity to the Fermi level, while the rest are energetically well separated. The zoom-in band structure around Fermi level [Fig.~\ref{fig1}(c)] reveals a band gap induced by the in-plane exchange field. This small-gap region surrounding K point originates from a nodal line protected by the horizontal mirror symmetry $\mathcal{M}_z$ in the absence of in-plane magnetization. The eigenvalues of the two central bands near the Fermi level are opposite: $+i$ and $-i$, which mean no hybridization of the two bands. The introduction of in-plane magnetization along the $x$ direction disrupts the $\mathcal{M}_z$ symmetry, rendering the nodal line gaped. The small gap between the two central bands and the large separation with other bands make this model suitable for demonstrating the dominant FSM spin response.

The magnetic point group of this model is $m^\prime m2^\prime$, which enables three independent components of the  intrinsic nonlinear spin response tensor: $\alpha_{xxy}$, $\alpha_{yxx}$, and $\alpha_{yyy}$. For illustration, we plot $\alpha_{xxy}$
in Fig.~\ref{fig2}(d) as a function of chemical potential $\mu$. Here, we evaluate both the full  $\alpha_{xxy}$  in Eq.~(\ref{total}) and  $\alpha_{xxy}^\text{QM}$ in Eq.~(\ref{FSM}). One can see that first, both results are peaked around $\mu=0$, showing that the nonlinear response is dramatically enhanced by the nodal loop. Second, the result of $\alpha_{xxy}^\text{QM}$ shows negligible difference from the full result, confirming our claim (\ref{dom}) that the intrinsic response is dominated by the quantum metric term.

%by using the complete BCP expression [Eq. (\ref{total})], its FSM contribution [Eq. (\ref{FSM})], and the two-band FSM contribution [Eq. (\ref{FSM-twoband})]. As shown in Fig.~\ref{fig1}(d), we find that the Fubini-Study quantum metric contribution dominates overwhelmingly the intrinsic nonlinear response. The two-band approximation of Fubini-Study metric is also perfect (thus not shown).

To make further connection between quantum metrics and nodal lines, we plot in Fig.~\ref{fig2}(e,f) the momentum-space distribution of $g_{xy}$ and $\mathfrak{g}_{xy}$ for the top valence band, which are involved in $\alpha_{xxy}^\text{QM}$.
One sees that these metric components are prominently concentrated along the two gaped nodal loops (marked by the dashed curves in the figure). This demonstrates our proposal that band (anti)crossings, especially nodal lines, are the desired band features to amplify the quantum metric tensor and nonlinear responses.

\textcolor{blue}{\emph{Application to 2D CrSBr.}} The model study has confirmed our general idea. The next question is whether such band topology enhanced nonlinear response can be appreciable in real materials. We answer this question by studying a concrete material, monolayer CrSBr.

\begin{figure}
	\includegraphics[width=8.6cm]{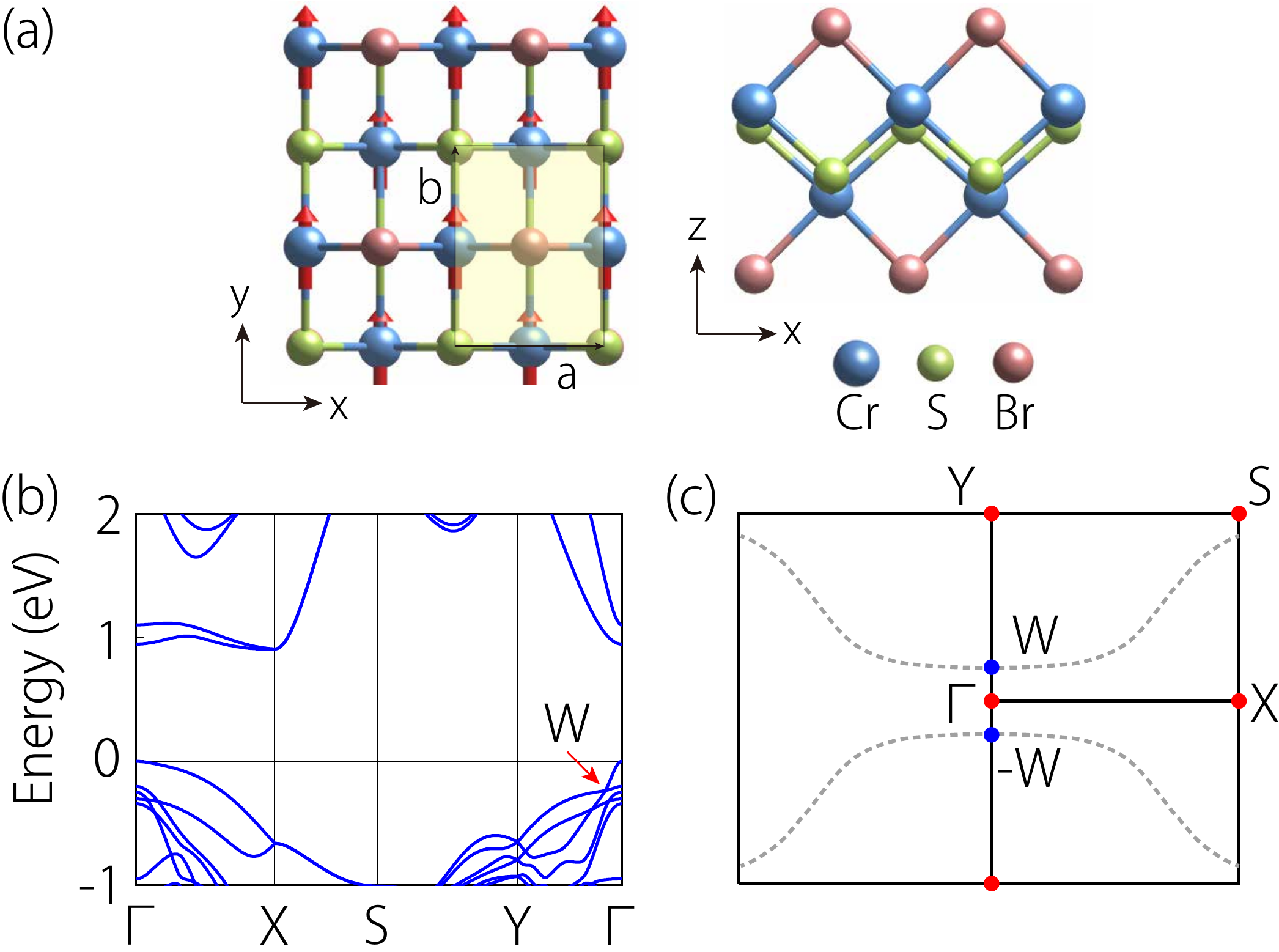}
	\caption{(a) Top and side views of monolayer CrSBr. The shaded rectangle indicates the unit cell. The red arrows denote the local spins on Cr sites in ground state. (b) Calculated band structure of monolayer CrSBr. (c) Brilliouin zone. The dashed lines indicate the two gaped nodal lines. $\pm W$ are the two Weyl points. 
		\label{fig3}}
\end{figure}

Monolayer CrSBr is a newly realized 2D ferromagnetic semiconductor \cite{guo2018chromium, yang2021triaxial, wilson2021interlayer, lee2021magnetic, telford2022coupling, rizzo2022visualizing, boix2022probing, wang2023magnetically, dirnberger2023magneto}, with a Curie temperature $\sim$146 K \cite{lee2021magnetic}. As depicted in Fig.~\ref{fig3}(a), it has an orthorhombic lattice structure with $D_{2h}$ point group symmetry. Our calculations based on density functional theory (DFT) yield lattice constants of $a = 3.54$ \AA\ and $b = 4.73$ \AA, consistent with previous results \cite{yang2021triaxial,lee2021magnetic} (calculation details are given in \cite{supp}). Experiments showed that the magnetic easy axis is along the $y$ axis [Fig.~\ref{fig3}(a)]. The corresponding magnetic point group is $m^{\prime}mm^{\prime}$, which contains the inversion symmetry. It follows that the linear spin response is forbidden, and we have to consider the nonlinear response. Among the symmetry allowed components of $\alpha_{abc}$,  $\alpha_{xxy}$ can exert a torque on the equilibrium magnetization, so we focus on this component in the following.

\begin{figure}
	\includegraphics[width=8.6cm]{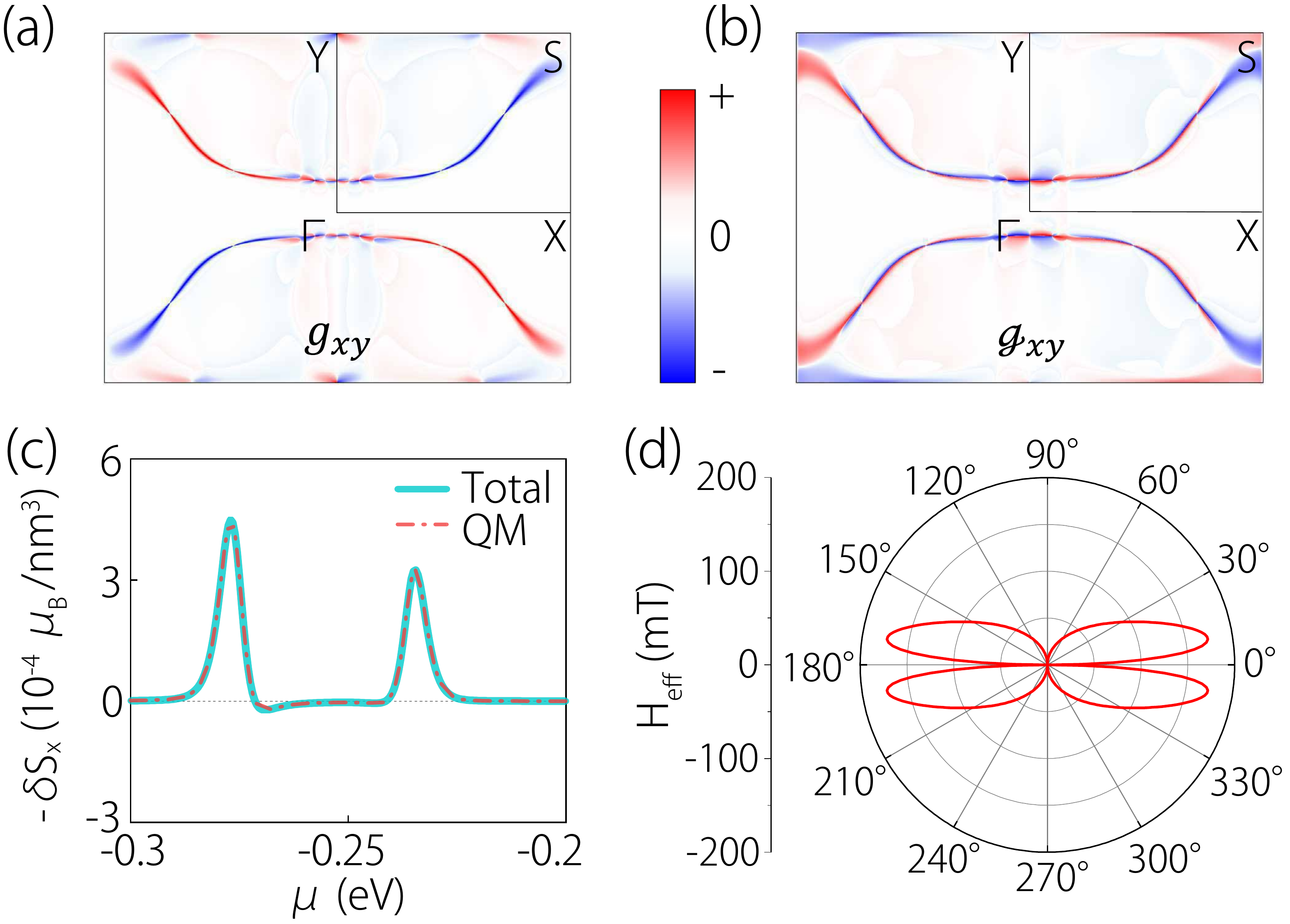}
	\caption{(a) and (b) show the distribution of quantum metric $g_{xy}$ and $\mathfrak{g}_{xy}$ of the top valence band in Brillouin zone. (c) Calculated nonlinear spin polarization versus chemical potential, under a moderate field $E = 10^5~\text{V/m}$ which makes an angle $\theta = 45^{\circ}$ with the $x$ axis. Temperature is taken at 20 K.  The total result (cyan solid curve) is almost entirely from the quantum metric contribution (red dash-dotted curve). (d) Angular dependence of current induced effective magnetic field due to QMT, under a driving current density of $j=10^7$ A/cm$^2$.
		\label{fig4}}
\end{figure}

The calculated band structure of monolayer CrSBr is plotted in Fig.~\ref{fig3}(b). One notices that there is a Weyl point $W$ on $\Gamma$-$\text{Y}$ path near the valence band top (marked by the arrow). This point is protected by the $C_{2y}$ symmetry.
Further inspection of the band structure uncovers that this Weyl point actually belongs to a nodal line traversing the Brillouin zone along the $x$ direction (see, e.g., Fig.~\ref{fig3}(c)). Owing to the $\mathcal{M}_y$ mirror, there are in fact a pair of such nodal lines. The two lines are gaped slightly by spin-orbit coupling (maximal local gap $\sim$ 13 meV) except at the Weyl points.
In Figs.~\ref{fig4}(a) and ~\ref{fig4}(b), we plot the distribution of $g_{xy}$ and $\mathfrak{g}_{xy}$ of the top valence band obtained from DFT calculation, which confirms that their values are peaked around the two nodal lines.

%connecting the highest two valence bands in the absence of spin-orbit coupling. This nodal line is protected by spin rotational symmetry (detailed analysis in \cite{supp}) \cite{feng2019discovery} and is gaped slightly by spin-orbit coupling, except for the Weyl point, enabling sizable topological spin responses. In Figs.~\ref{fig2}(c) and (d), we plot the distribution of $k$-space FSM $g_{xy}$ and $k$-$m$ space FSM $\mathfrak{g}_{xy}$ of the top valence band in the Brillouin zone. One indeed sees that both metrics mainly concentrate at the nodal-line region.

Next, we evaluate both $\alpha_{xxy}$ and  $\alpha_{xxy}^\text{QM}$ by combining our theory with DFT calculations.
Figure~\ref{fig4}(c) shows induced nonlinear spin polarization $\delta S_x$ versus $\mu$ under a moderate driving field $E = 10^5~\text{V/m}$ \cite{Jungwirth2018,Song2019} along the direction $(\hat{x}+\hat{y})/\sqrt{2}$. We have the following observations. (1) The curves obtained from  $\alpha_{xxy}$ and  $\alpha_{xxy}^\text{QM}$ exhibit negligible difference, meaning that the response is almost entirely from the quantum metric contribution. Thus, it is justified to refer to the resulting torque here as QMT. (2) The peak of the response reflects the enhancement from the topological band crossing. The upper peak at $\mu = -0.23$ eV corresponds to the contribution from the nodal line around points $W$ and $-W$, while the lower peak at $\mu = -0.28$ eV arises from the small gap regions at $(k_x, k_y) = (\pm 0.083, \pm 0.115)$.
(3) The magnitude is on the order of $10^{-5}$ $\mu_{B}$/nm$^3$, and the peak value (upper peak) can reach $\sim 3.2 \times 10^{-4}$ $\mu_{B}$/nm$^3$, which is very significant and much larger than the previously measured linear spin polarization in noncentrosymmetric (Ga,Mn)As ($10^{-9}$ to $10^{-6}$ $\mu_{B}$/nm$^3$ \cite{Geller2009,Vyborny2011,Kurebayashi2014})

The induced spin polarization exerts a torque $\bm T=\bm m\times \bm H^\text{eff}$ on the magnetization, with the effective field
\begin{equation}
  H^{\mathrm{eff}}_a = -(J_{ex}/g\mu_{B})\delta S_a,
\end{equation}
where $J_{ex}$ is the exchange coupling between carrier spin and local moment, and $g$ is the spin g factor. The magnitude of this effective field is an important factor used for analyzing magnetic dynamics \cite{Manchon2019}. In Fig.~\ref{fig4}(d), we plot the variation of $H^{\mathrm{eff}}_x$ versus the in-plane direction of driving current density $j$, with $j=10^7$ A/cm$^2$. We find that in a wide angle range, the torque efficiency $\zeta=H^{\mathrm{eff}}_x/j$ is on the order of 100 $\text{mT}$ per $10^7 \text{A/cm}^2$. The maximal $\zeta$ can reach about 173 $\text{mT}$ per $10^7 \text{A/cm}^2$, which is at least an order of magnitude larger than usual values of linear spin-orbit torques (0.1 to 10 $\text{mT}$ per $10^7 \text{A/cm}^2$) found in inversion asymmetric systems, such as (Ga,Mn)As~\cite{Vyborny2011}, NiMnSb~\cite{Jungwirth2016}, and various multilayer structures~\cite{Manchon2019}.

Is this QMT sufficient for magnetic switching? Usually, this is judged by comparing the effective field $H^\text{eff}$ with the magnetic anisotropy field $H_K$. Consider a driving current density of $5 \times 10^7\ \text{A/cm}^2$. This can generate an effective field $H^\text{eff}\sim 4.3$ T, which is already much larger than the anisotropy field for monolayer CrSBr ($\sim 1$ to 2 T) \cite{boix2022probing}. Furthermore, it is noted that the QMT is a (anti)damping-like torque \cite{Manchon2019}, i.e., it acts in the form of
$\Hat{\bm m}\times (\Hat{\bm m}\times \Hat{ y})$ in the switching process. It has been recognized that  (anti)damping-like torques are more efficient for switching, as the anisotropy field that it competes with will be further multiplied by the Gilbert damping factor $\eta$ and $\eta$ is usually a small number $\ll 1$ \cite{Manchon2019}. Here, if we take $\eta=0.01$ as a typical value for Cr-based 2D materials, $\eta H_K$ in CrSBr is only about 10 to 20 mT. Therefore, it is highly possible to achieve full-electric magnetic switching via QMT in this system.

%makes it possible for the first time to envisage magnetization switching by nonlinear spin response. For example, an input current density of $5 \times 10^7 \text{A/cm}^2$ can trigger an effective field of 4.3 $\text{T}$, which is much larger than the magnetic anisotropy field $H_K$ about 1 $\sim$ 2 T in this material \cite{boix2022probing}. Moreover, the quantum metric torque in CrSBr is a damping-like torque \cite{Manchon2019}. Initially, the nonequilibrium spin polarization $\delta S_x$ tilts the equilibrium magnetization in the easy ($y$) axis towards the $z$ direction, and the direction of the rotated
%equilibrium magnetization is denoted by a unit vector $\Hat{\bm n}$ (see the schematics in Fig.~\ref{fig2}(g)). It is then apparent that the intrinsic nonlinear torque takes the form of $\Hat{\bm n}\times (\Hat{\bm n}\times \Hat{\bm y})$, which competes with the magnetic anisotropy field multiplied by the Gilbert damping factor $\eta$ \cite{Manchon2019}. Given the fact that $\eta$ is usually much less then 1 as well as its typical value of 0.01 in Cr-based 2D materials \cite{gilmore2007identification, hiramatsu2018first, esteras2022magnon}, $\eta H_K$ in CrSBr is only about 10 $\sim$ 20 mT. Therefore, it is possible that the quantum metric torque driven by input current density of $5 \times 10^6 \text{A/cm}^2$ can lead to switching.

\textcolor{blue}{\emph{Discussion.}}--We have unveiled the critical role of quantum metric in nonlinear spintronic responses. Our calculations demonstrate that the intrinsic nonlinear spin-orbit torque induced by topological band structures can be  almost entirely attributed to the quantum metric contribution.
%This refreshed understanding inspires the strategy to magnify the response by resorting to topological nodal-line metals.
And our evaluation for 2D CrSBr points out for the first time the possibility of magnetic switching by QMT.
Our result also suggests that the nonlinear spintronic responses offer a new route to probe quantum metric as well as topological band structures.

We have focused on the intrinsic spin response. There also exist extrinsic responses related to field driven off-equilibrium distribution function, which depend on scattering \cite{Xiao2023NLSOT}. For CrSBr, we have also estimated the extrinsic responses, which are found to be two to three orders of magnitude smaller than the QMT \cite{supp}. This hence provides a good opportunity to probe the intrinsic effect and quantum metric.

%For comparison, we also calculate the nonlinear spin response related to field driven off-equilibrium distribution function, which can be at the first or the second order of relaxation time \cite{Xiao2023NLSOT}. Both these contributions peak when the chemical potential is located around band near-degeneracies, but are two to three orders of magnitude smaller than the intrinsic quantum metric effect.
%%For instance, at the chemical potential of $\mu = -0.23$ eV, they are at the order of $10^{-7}$ $\mu_{B}$/nm$^3$ and $10^{-6}$ $\mu_{B}$/nm$^3$.
%Therefore, in this material, the quantum metric contribution appears as the dominant factor in the overall nonlinear spin response.
%
%We have uncovered that a quantum metric nonlinear spin-orbit torque can emerge in topological band structures, which is amplified significantly by such bands and hence dominates the second order nonlinear torque. The two-band character of topological structures enables evaluating the quantum metric torque by simply counting the interband coherence between the two central topological bands, even if multiple bands may appear around the Fermi level. Conducting first-principles calculations to monolayer CrSBr, we find giant quantum metric torque, which may lead to magnetization reversal by nonlinear effect for the first time.

The possibility of magnetic switching by QMT establishes a close link between nonlinear spintronics and topological materials. 
%Our work thus opens the door to the research field of topological nonlinear spintronics and demonstrates its promise in magnetic memory application. 
The proposed strategy will facilitate the search of new spintronics platforms with large quantum metric response enabled by band topology.
In particular, it would be desired to have multiple nodal points/lines at similar energies such that quantum metric contributions can be concentrated in a narrow energy window to deliver a large response.

%with less dispersive nodal line or higher-dimensional nodal structures, such as Co$_{2}$MnGa \cite{Chang2017,sakai2018giant,Belopolski2019discovery} and Fe$_{3}$Sn \cite{chen2022large}, are more favored for stronger effect because of the larger density of states.
%
%
%
%topological nonlinear spintronics and demonstrates its promise in magnetic memory applications. The results may facilitate the search for other magnets with topological bands that can support even larger quantum metric torque. For instance, the recently proposed ferromagnets Gd$_2$C \cite{liu2020ferromagnetic}, LaCl \cite{liu2018intrinsic}, and Fe$_4$GeTe$_2$ \cite{bera2023anomalous} are potential candidates due to nodal line characteristic in their band structures. In particular, magnets with less dispersive nodal line or higher-dimensional nodal structures, such as Co$_{2}$MnGa \cite{Chang2017,sakai2018giant,Belopolski2019discovery} and Fe$_{3}$Sn \cite{chen2022large}, are more favored for stronger effect because of the larger density of states.

\bibliographystyle{apsrev4-2}
\bibliography{intNSME_ref.bib}
\end{document}